# Axion Insulator State in Ferromagnetically Ordered CrI$_3$/Bi$_2$Se$_3$/MnBi$_2$Se$_4$ Heterostructure

Y. S. Hou[1], J. W. Kim[2] and R. Q. Wu[1]

[1] Department of Physics and Astronomy, University of California, Irvine, CA 92697-4575, USA

[2] Department of Physics, Incheon National University, Incheon 22012, Korea

## Abstract

Ferromagnetic (FM) axion insulators may greatly simplify experimental explorations of the topological magnetoelectric effect, as the FM ordering can be stable in a broad range of magnetic field. Through density functional theory calculations and four-band model studies, we find that two-dimensional van der Waals FM CrI$_3$ and MnBi$_2$Se$_4$ overlayers induce opposite-sign exchange fields in the topological surface states of Bi$_2$Se$_3$ film when they are ferromagnetically ordered. Consequently, CrI$_3$/Bi$_2$Se$_3$/MnBi$_2$Se$_4$ heterostructure is a robust FM axion insulator, according to surface-resolved Berry curvature calculations. Our work opens up new opportunities for the design of materials with the topological magnetoelectric effect.

Email: wur@uci.edu

The topological magnetoelectric (TME) effect [1-4], a quantized version of the magnetoelectric effect [5], is the hallmark response of the magnetized three-dimensional (3D) topological insulators (TIs) to the applied electric and magnetic fields **E** and **B**. This intriguing effect shows up in the axion insulator (AI) phase [6-18], with an effective action $S_\theta$ in the Maxwell's equations [1]

$$S_\theta = \frac{\alpha}{2\pi} \int d^3x dt \frac{\theta}{2\pi} \mathbf{E} \cdot \mathbf{B} \quad (1)$$

where $\alpha = e^2/\hbar c$ is the fundamental fine structure constant, and $\theta = 0$ or $\pi$ for the topologically trivial and nontrivial cases [19], respectively. One important consequence of the TME effect in AIs is the cross induction between electric and magnetic properties as [1]:

$$\mathbf{P} = \alpha/4\pi \, \mathbf{B} \quad (2),$$

$$\mathbf{M} = -\alpha/4\pi \, \mathbf{E} \quad (3).$$

where **P** and **M** are the electric polarization and magnetization, respectively. Obviously, studies of AIs may not only advance fundamental physics, but also open a new vista of technological innovations, particularly for quantum devices [20]. To date, experimental investigations of AIs have focused on FMI/TI/FMI heterostructures [6-8,21], with FMI standing for a ferromagnetic (FM) insulator. To enter the AI phase, an antiferromagnetic (AFM) ordering between the two FMIs is typically required [1,22], which only survives in a narrow range of external magnetic field $B$, i.e., $B_{1c} < |B| < B_{2c}$ [7,8,22]. These AFMAI states have two major disadvantages for the experimental realization of TME effect, even worse, for the devise applications. First, the difference between the coercive fields of the two FMIs, $|B_{2c} - B_{1c}|$, is typically small and hard to control. Second, the constant need of a sizable external magnetic field in operations may consume energy and makes it difficult to implement in practical devices. To circumvent these problems, it is desired to realize the AI state with a FM ordering (FMAI) between two the FMIs, which can be easily established with a small or even zero external magnetic field.

It is interesting to compare the performances of between FMAIs and AFMAIs when one sweeps magnetic field. The AFM ordering between the two FMIs survives in the range $B_{1c} < |B| < B_{2c}$ while the FM ordering appears in all range of $B$; see the evolutions of the magnetizations in Fig. 1a and 1b. Consequently, the AFMAI only exists in a range $B_{1c} < |B| < B_{2c}$ (highlighted by cyan in Fig. 1a) whereas the system is a Chern insulator in other regions of $B$. In contrast, the FMAI appears in a broad range of $B$ except the yellow regions in Fig. 1b and thus is much easier to be established, especially when $|B_{2c} - B_{1c}|$ is small. Furthermore, the FM ordering can be reached in high magnetic field regardless the exchange interaction between the two FMIs. A large TME signal (e.g., **P**) can be thereby obtained according to Eq. (2). The possibility of having the TME effect at either low or high $B$ makes the FMAIs very attractive for fundamental studies and applications.

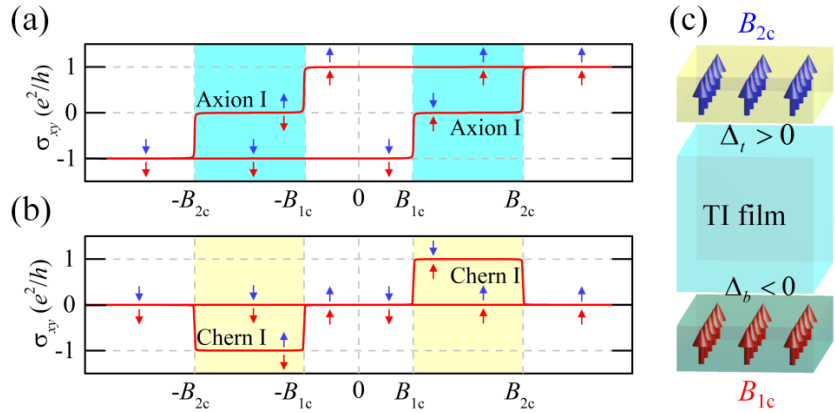

*Figure 1 (color online) The schematic dependence of AHC versus applied magnetic field in (a) AFMAI and (b) FMAI. The magnetizations of the top and bottom FMIs are sketched by the blue and red arrows, respectively. (c) The design of FMAI with a vdW-FMI/TI/vdW-FMI heterostructure. $B_{1c}$ ($B_{2c}$) is the coercive field of the bottom (top) FMI.*

In principle, the AI phase only requires that the anomalous Hall conductances (AHCs) at two interfaces have opposite signs [1,22]. Because two identical FMIs induce the same exchange splitting in the topological surface states (TSSs) of a TI film when they are ferromagnetically ordered, one needs to utilize two different FMIs to produce the FMAI state. Recent discoveries of many van der Waals (vdW) FMIs [23-25] offer a large pool of materials for fulfilling this purpose. As one places two-dimensional (2D) vdW FMIs on both sides of 3D TI films (vdW-FMI/TI/vdW-FMI, Fig. 1c), TSSs of the TI film are uniformly magnetized and the spin polarized band gaps of these systems are typically large and right at the Fermi level [26,27]. Most 2D vdW FMIs such as $CrI_3$ monolayer (ML) have reasonably high Curie temperatures [24], so it is conceivable that the TME effect can be conveniently observed in lab conditions. Obviously, the major challenge for the realization of FMAIs is to find appropriate materials combinations that have 1) robust spin polarized band gaps for TSSs, 2) opposite AHCs at two interfaces, and 3) adequately high structural and magnetic stabilities.

In this Letter, we demonstrate that the heterostructure of $CrI_3$ ML, $MnBi_2Se_4$ (MBS) septuple-layer (SL) and 3D TI $Bi_2Se_3$ (BS) film ($CrI_3$/BS/MBS) is an excellent FMAI. Based on the density functional theory (DFT) calculations and four-band model analyses, we find that $CrI_3$ ML and MBS SL in $CrI_3$/BS/MBS provide opposite exchange fields to the TSSs of BS film and produce opposite AHCs at the top and bottom interfaces when their magnetizations are parallel to each other. This establishes a robust FMAI state in a large range of external magnetic field, and $CrI_3$/BS/MBS is thus a very promising system for the use in nanodevices.

Our DFT calculations are performed using the Vienna *Ab initio* Simulation Package with the generalized gradient approximation [28,29]. We use the projector-augmented wave pseudopotentials to describe the core-valence interaction [30,31] and an energy cutoff of 500 eV for the basis expansion. To take account the strong correlation effect among Cr and Mn 3*d* electrons, we utilize $U_{Cr} = 3.0$ eV and $J_{Cr}^{H} = 0.9$ eV for Cr [32] and $U_{Mn} = 6.0$ eV and $J_{Mn}^{H} = 1.0$ eV for Mn [33] in the self-consistent calculations. The in-plane lattice constant of BS films, $a_{BS}$=4.164 Å, is adopted for all $CrI_3$/BS/MBS heterostructures. The atomic alignments at $CrI_3$/BS and MBS/BS interfaces are fully optimized, as done in our previous studies [10,32] and also given in Fig. S1 of Supplementary materials (SM) [34]. The vdW interaction is described by the nonlocal vdW functional (optB86b-vdW) [35,36]. For the convenience of discussions below, the *z*-axis is perpendicular to the BS film, pointing from the MBS SL to the $CrI_3$ ML, and the *x*- and *y*-axes are in the surface plane.

In a vdW-FMI/TI/vdW-FMI heterostructure with a thick TI film, the FMAI state with a quantized TME effect is expected if the top ($\Delta_t$) and bottom ($\Delta_b$) exchange fields have opposite signs [1,22], i.e., $\Delta_t \Delta_b < 0$, as depicted in Fig. 1c. Here, CrI$_3$ ML and MBS SL are chosen as vdW-FMIs that have been actively studied in recent years to see if they may produce opposite exchange fields to TSSs of BS film when they are ferromagnetically ordered. CrI$_3$ ML is found to induce a sizable nontrivial spin gap for the TSSs of BS film [32]. A recent experiment [26] demonstrated those: 1) MBS SL consists of Se-Bi-Se-Mn-Se-Bi-Se atom layers (Fig. S2 in SM); 2) its Curie temperature is 300 K; and 3) it opens a large gap at the TSSs of BS film. Moreover, the magnetic ions in CrI$_3$ ML and MBS SL have different electronic configurations in their 3$d$ shells, i.e., $t_{2g}^{3\uparrow} e_g^{0\uparrow} t_{2g}^{0\downarrow} e_g^{0\downarrow}$ for Cr$^{3+}$ and $t_{2g}^{3\uparrow} e_g^{2\uparrow} t_{2g}^{0\downarrow} e_g^{0\downarrow}$ for Mn$^{2+}$. In this case, the top and bottom surfaces may have opposite half-integer AHCs, i.e., $\sigma_{xy}^t = \text{sgn}(\Delta_t)(e^2/2h)$ and $\sigma_{xy}^b = \text{sgn}(\Delta_b)(e^2/2h)$.

We first investigate the band structure of CrI$_3$/6QL-BS/MBS with a FM ordering between CrI$_3$ ML and MBS SL. Unless otherwise noted, the FM ordering with Cr$^{3+}$ and Mn$^{2+}$ ions' magnetic moments pointing along the $z$ axis (Fig. 2b) is assumed in our calculations. As a result, CrI$_3$/6QL-BS/MBS is insulating with a sizable band gap of 5.6 meV. By projecting the wave functions of bands to the BS film, we find that the Dirac cone features of its TSSs are reserved (Fig. 2a). Furthermore, the top and bottom surface states are distinctly separated in energy (Figs. 2c and 2d). For the convenience of discussions, we number the four bands near the Fermi level according to their energy sequences (Fig. 2c). Near the Γ point, bands II (I) and III (IV) belong to the top (bottom) surface. Gaps opening at the cross points of band I and band II (marked as black asterisks in Figs. 2c and 2d, also shown in Fig. S3 in SM) are tiny, suggesting an extremely weak couplings between the top and bottom surfaces.

The combination of the insulating nature and FM ordering in CrI$_3$/6QL-BS/MBS can result in three different quantum phases, i.e., Chern insulator, FMAI and topologically trivial insulator. To distinguish them, we calculate AHC, $\sigma_{xy}$, using Wannier90 package [37]. We observe from Fig. 2e that $\sigma_{xy}$ is zero when the Fermi level locates in the gap. This first excludes that CrI$_3$/6QL-BS/MBS is a Chern insulator. However, because both FMAI and topologically trivial insulator have a zero $\sigma_{xy}$, it is not possible to distinguish these phases at this stage.

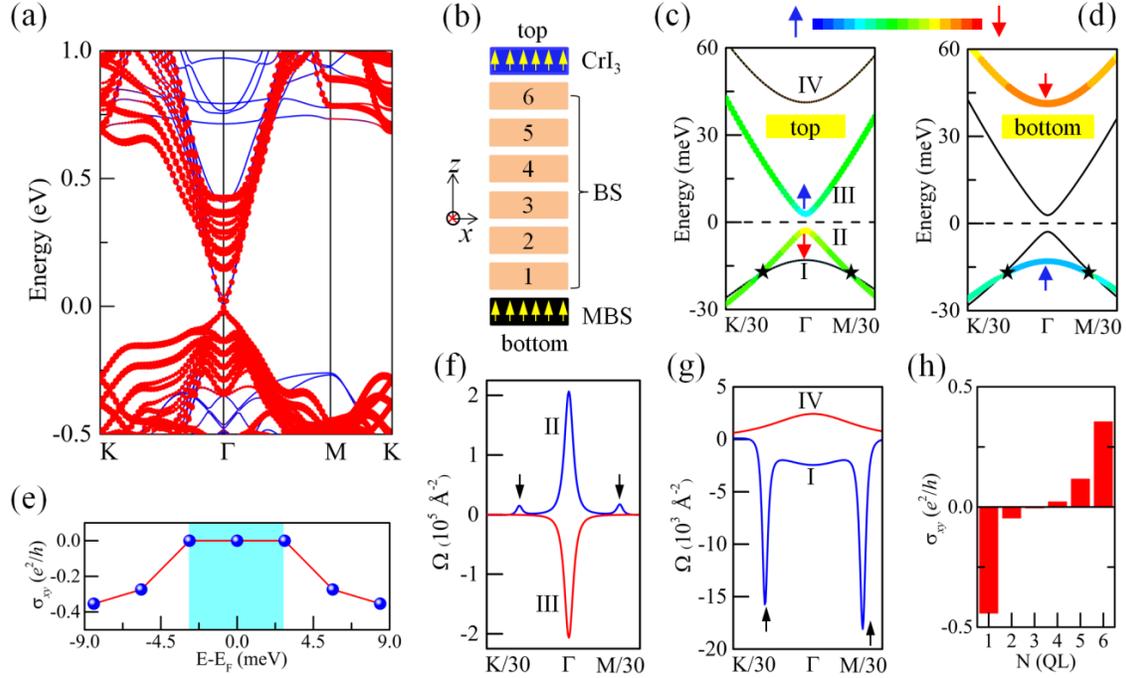

*Figure 2 (color online) (a) BS-projected (red dots) band structure of FM CrI$_3$/6QL-BS/MBS. (b) The FM ordering and the indices of the quintuple layer (QL) of BS film in FM CrI$_3$/6QL-BS/MBS. (c) and (d) show the spin components of the top and bottom surface states. The color bar shows the spin weight. Arrows depict the dominant spin components at the Γ point. (e) The dependence of AHC on the position of the Fermi level. (f) and (g) show the band-resolved Berry curvatures of the four bands near the Fermi level. Roman numerals in (f) and (g) follow the band indices in (c). (h) The QL-resolved AHCs of FM CrI$_3$/6QL-BS/MBS and the indices of QL, N, follow those in (b).*

To shed light on the signs of the top- and bottom-surface exchange fields in CrI$_3$/6QL-BS/MBS, we examine the spin components of bands I-IV. For bands II and III at the top surface, there is a strong spin intermixing and band II has a large weight in the spin-down channel (Fig. 2c). For bands I and IV at the bottom surface, the low-lying band I has a dominant spin-up component (Fig. 2d). These suggest that the top and bottom FMIs have opposite exchange fields on the TSSs of the BS film, namely, $\Delta_t > 0$ and $\Delta_b < 0$. To further verify this, we calculate the band-resolved Berry curvatures using Wannier90 package [37]. As shown in Figs. 2f and 2g, band II and band III of the top surface have positive and negative Berry curvatures, whereas band I and band IV of the bottom surface have negative and positive Berry curvatures, respectively. Besides, the opposite Berry curvatures of the top and bottom surfaces are clearly demonstrated by the QL-resolved AHCs (Fig. 2h) and the spatially resolved Berry curvatures (Fig. S4 in SM) [38]. They are all consistent with conjecture of $\Delta_t > 0$ and $\Delta_b < 0$, as desired for the realization of FMAI. Furthermore, we find that this FMAI state is robust against of the choices of $U_{Cr}$, $U_{Mn}$ (Fig. S5 and S6 in SM) and different DFT functional (Fig. S7 in SM).

Another way to check if a system is a FMAI is through its complementary feature. As indicated in Fig. 1b, systems enter the Chern insulator state when the two FMIs are antiferromagnetically ordered. To show this for $CrI_3$/6QL-BS/MBS, we set an AFM ordering between $CrI_3$ ML and MBS SL (Fig. S8 in SM) and examine its topological properties. As expected, Chern number $C_N=-1$ is obtained, with a quantized AHC $\sigma_{xy} = -e^2/h$ (Fig. S8 in SM) for $CrI_3$/6QL-BS/MBS with an AFM ordering between the two FMIs $CrI_3$ ML and MBS SL.

Note that the couplings between the top and bottom surface states lead to two extra side peaks (indicated by the black arrows in Fig. 2f and 2g) in the band-resolved Berry curvatures around the cross points of bands I and II. This should be avoided as the coupling between the top and bottom surface states, though it is very weak, is unwanted in AIs for achieving the quantized TME effect [1,22]. As direct DFT calculations for topological properties of large systems are still too expensive, we adopt a low-energy effective four-band model, to investigate the dependences of $\Delta_t$, $\Delta_b$, $\Delta_h$ and Berry curvatures on the thickness of the BS film in $CrI_3$/BS/MBS. With the bases of $\{|t,\uparrow\rangle, |t,\downarrow\rangle, |b,\uparrow\rangle, |b,\downarrow\rangle\}$, the effective Hamiltonian is [22,39-42]

$$H(k_x,k_y)=Ak^2+\begin{bmatrix} v_F(k_y\sigma_x-k_x\sigma_y) & M_k\sigma_0 \\ M_k\sigma_0 & -v_F(k_y\sigma_x-k_x\sigma_y) \end{bmatrix}+\begin{bmatrix} \Delta_t\sigma_z & 0 \\ 0 & \Delta_b\sigma_z \end{bmatrix}+\begin{bmatrix} V_P\sigma_0 & 0 \\ 0 & -V_P\sigma_0 \end{bmatrix} \quad (4).$$

Here, ↑ and ↓ denote the spin-up and spin-down states; $v_F$ is the Fermi velocity; $k_x$, $k_y$ and $k^2=k_x^2+k_y^2$ are in-plane wave vectors; $\sigma_{x,y,z}$ ($\sigma_0$) is Pauli (2-by-2 identity) matrix. $M_k=\Delta_h+Bk^2$ describes the coupling between the top and bottom surface states and $\Delta_h$ is the coupling induced gap. Since $CrI_3$ ML and MBS SL are two different FMIs, there is a nonvanishing inversion asymmetry potential (IAP), $V_P$ [39,43-45]. The formulas in Ref. [46] are used to calculate Berry curvature $\Omega$ and Chern number $C_N$.

By fitting the DFT band structure to the four-band model [Eq. (4)], we obtain the fitted $\Delta_h=0.8$ meV, $\Delta_t=2.9$ meV, $\Delta_b=-26.9$ meV and $2|V_P|=13.2$ meV for $CrI_3$/6QL-BS/MB. As shown in Fig. S9 in SM, the high quality of reproductions of bands from DFT and Berry curvatures from Wannier90 by the four-band model indicates that the four bases are adequate to describe the topological properties of states near the Fermi level for $CrI_3$/BS/MBS. We see that either the top-surface exchange field $\Delta_t$ or the bottom-surface exchange field $\Delta_b$ are obviously larger than $\Delta_h$. By decomposing the Berry curvatures to the top and bottom contributions in the frame of the four-band model, we obtain nonzero and opposite top- and bottom-surface AHCs (Fig. S10 in SM). Therefore, 6QL-BS is thick enough for the realization of the FMAI state.

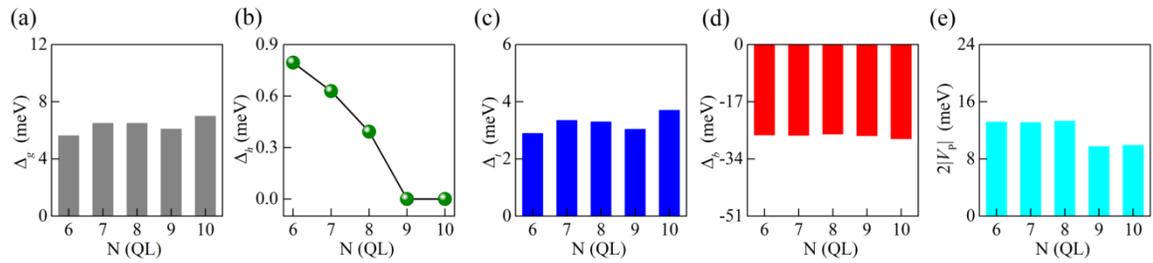

Figure 3 (color online) Dependences of (a) the DFT calculated global gap $\Delta_g$, (b) the coupling induced gap $\Delta_h$, (c) the top-surface exchange field $\Delta_t$, (d) the bottom-surface exchange field $\Delta_b$, and (e) the IAP $2|V_P|$ on the thickness of BS film in FM $CrI_3$/BS/MBS.

Now we investigate the effect of the thickness of BS film on the FMAI state in CrI$_3$/BS/MBS by varying BS film from six to ten QLs. All CrI$_3$/BS/MBS are insulating with a global band gap of about 6 meV (Fig. 3a). As the thickness of BS film increases, $\Delta_h$ gradually decreases and becomes zero at the ten-QL BS film (Fig. 3b). Values of $\Delta_t$, $\Delta_b$, and $V_P$ depend weakly on the thickness of BS film (Fig. 3c-3e). Overall, it is plausible that the FMAI state persists in CrI$_3$/BS/MBS as long as the BS film is thicker than 6 QLs. Especially, CrI$_3$/10QL-BS/MBS has opposite half-integer AHCs at its top and bottom surfaces (Fig. S11 in SM), so it exhibits the exactly quantized TME effect.

Furthermore, we elucidate the effect of $V_P$ on the FMAI state in vdW-FMI/TI/vdW-FMI, because it could be comparable to the exchange field, as in the case of CrI$_3$/BS interface. A nonvanishing $V_P$ is caused either intrinsically by the structure inversion asymmetry due to the use of two different vdW-FMIs or extrinsically by the operating bias in applications [47]. For simplicity, we consider the case of a thick BS film, i.e., $M_k$=0 in Eq. (4). In this case, the dispersions of the four-band model [Eq. (4)] are as follows:

$$\varepsilon_\pm^t(k_x, k_y) = Ak^2 \pm \sqrt{\Delta_t^2 + v_F^2 k^2} + V_P \qquad (5),$$

$$\varepsilon_\pm^b(k_x, k_y) = Ak^2 \pm \sqrt{\Delta_b^2 + v_F^2 k^2} - V_P \qquad (6).$$

Eq. (5) and (6) indicate that the presence of $V_P$ shifts the bands of the top and bottom surfaces in opposite ways. Due to this shift, vdW-FMI/TI/vdW-FMI is in the FMAI state (the metallic state) only when $2|V_P|$ is smaller (larger) than the critical point $|\Delta_t|+|\Delta_b|$ (Fig. 4). Besides, $V_P$ reduces the gap of the FMAI state linearly when $2|V_P| > \left\||\Delta_t|-|\Delta_b|\right\|$, but it has no effect on the FMAI state when $2|V_P| < \left\||\Delta_t|-|\Delta_b|\right\|$. Therefore, 2D vdW FMIs that produce a large exchange field on the TSSs are advantageous to obtain the FMAI state. To show the importance of $V_P$ on the FMAI state, we take the ferromagnetically ordered CrI$_3$/10QL-BS/EuBi$_2$Se$_4$ (EBS) as an example. The fitting parameters for this system are $\Delta_{t,2} = 3.2$ meV, $\Delta_{b,2} = -2.6$ meV, and $2|V_{P,2}|$=18.7 meV (Fig. S12 in SM) [34]. Since $|\Delta_{t,2}|+|\Delta_{b,2}| < 2|V_{P,2}|$, it should metallic according to the phase diagram (Fig. 4). Indeed, DFT band structures (Fig. S12 in SM) clearly show that CrI$_3$/BS/EBS with a FM ordering between CrI$_3$ ML and EBS SL is metallic, although its top and bottom surfaces have sizable and opposite exchange fields.

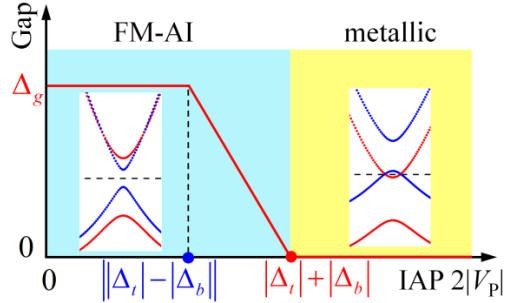

Figure 4 (color online) Phase diagram of the vdW-FMI/TI/vdW-FMI with a varying IAP 2|$V_P$|. Here, the global gap $\Delta_g$ is $\min(|\Delta_t|,|\Delta_b|)$. Insets depict the schematic band structures of FMAI and metallic states.

From the experimental perspectives, the fabrication of CrI$_3$/BS/MBS is feasible because 1) MBS SL has already been synthesized by δ-doping BS [26,48], and 2) CI$_3$/BS can be mechanically assembled owing to their vdW nature [49,50]. A practical scheme of growing of CrI$_3$/BS/MBS is sketched in Fig. S15 in SM. The MBS SL can be first synthesized on a suitable substrate following the methods of Ref. [26,48]. BS films can then be grown it through molecular-beam epitaxy [47]. Lastly, CrI$_3$ ML [24] can be put on BS/MBS to form the CrI$_3$/BS/MBS heterostructure.

To summarize, our systematic DFT calculations and four-band model studies indicate that CrI$_3$/BS/MBS may display the robust FMAI state. This arises from the fact that CrI$_3$ ML and MBS SL induce opposite-sign exchange fields into the TSSs of BS film when they are ferromagnetically ordered. Here we emphasize that this FMAI state is distinctly different from the recently studied AI state in the layered-AFM ordered topological insulator MnBi$_2$Te$_4$ [11,17,18] and Bi$_2$MnSe$_4$ [12]. From the band gaps of TSSs and Curie temperatures of the two FMIs, we perceive that the FMAI state in CrI$_3$/BS/MBS may survive up to a few tens Kelvins. While the realization of FMAI state is demonstrated with CrI$_3$/BS/MBS in this work, many other material choices may serve for the same purpose. For example, one may replace CrI$_3$ with Cr$_2$Ge$_2$Te$_6$ [25] and MBS with MnBi$_2$Te$_4$ [51]. As the family of the 2D vdW-FMIs grows rapidly [23], our findings point to a practical route for the realization of the quantized TME effect and should stimulate more experimental and theoretical investigations in this realm.

Work was supported by DOE-BES (Grant No. DE-FG02-05ER46237). DFT calculations were performed on parallel computers at NERSC supercomputer centers.